\documentclass[runningheads]{llncs}
\usepackage{graphicx}
\usepackage{mathrsfs}
\usepackage[colorlinks,linkcolor=blue]{hyperref}
\usepackage{makecell}
\usepackage{multirow}
\usepackage{cite}
\usepackage{amsmath,amssymb,amsfonts}
\usepackage{algorithmic}
\usepackage{graphicx}
\usepackage{textcomp}
\usepackage{subfigure}

\usepackage{xcolor}
\usepackage{booktabs}
\usepackage{diagbox}
\usepackage{tabularx}
\usepackage{setspace}
\usepackage{multirow}

\usepackage{cleveref}

\usepackage[misc]{ifsym} 

\def\BibTeX{{\rm B\kern-.05em{\sc i\kern-.025em b}\kern-.08em
		T\kern-.1667em\lower.7ex\hbox{E}\kern-.125emX}}

\begin{document}
\title{Style-invariant Cardiac Image Segmentation with Test-time Augmentation}
%
%
\author{Xiaoqiong Huang\inst{1,2}\thanks{Xiaoqiong Huang and Zejian Chen contribute equally to this work.}, Zejian Chen\inst{1,2\star}, Xin Yang\inst{1,2}, Zhendong Liu\inst{1,2}, Yuxin Zou\inst{1,2}, Mingyuan Luo\inst{1,2}, Wufeng Xue\inst{1,2}, Dong Ni\inst{1,2}\textsuperscript{(\Letter)}}

\authorrunning{Huang et al.}

\institute{
	\textsuperscript{$1$}School of Biomedical Engineering, Shenzhen University, Shenzhen, China\\
	\textsuperscript{$2$}Medical UltraSound Image Computing (MUSIC) Lab, Shenzhen University, China\\
	\email{nidong@szu.edu.cn} \\
}

\maketitle
\begin{abstract}
Deep models often suffer from severe performance drop due to the appearance shift in the real clinical setting. Most of the existing learning-based methods rely on images from multiple sites/vendors or even corresponding labels. However, collecting enough unknown data to robustly model segmentation cannot always hold since the complex appearance shift caused by imaging factors in daily application. In this paper, we propose a novel style-invariant method for cardiac image segmentation. Based on the zero-shot style transfer to remove appearance shift and test-time augmentation to explore diverse underlying anatomy, our proposed method is effective in combating the appearance shift. Our contribution is three-fold. First, inspired by the spirit of universal style transfer, we develop a zero-shot stylization for content images to generate stylized images that appearance similarity to the style images. Second, we build up a robust cardiac segmentation model based on the U-Net structure. Our framework mainly consists of two networks during testing: the ST network for removing appearance shift and the segmentation network. Third, we investigate test-time augmentation to explore transformed versions of the stylized image for prediction and the results are merged. Notably, our proposed framework is fully test-time adaptation. Experiment results demonstrate that our methods are promising and generic for generalizing deep segmentation models.
\end{abstract}

\section{Introduction}
Delineation of the left ventricular cavity (LV), myocardium (MYO), and right ventricle (RV) from cardiac magnetic resonance (CMR) images (multi-slice 2D cine MRI) is a common clinical task to establish the diagnosis. It is of great interest to develop an accurate automated segmentation method since manual segmentation is tedious and likely to suffer from inter-observer variability. Deep learning cardiac segmentation models have achieved remarkable success based on a large amount of labeled data. However, learning-based models often subject to severe performance drop due to testing data that has different distributions from the training data. This is a highly desirable but challenging task that makes deep models robust against the complex appearance shift of testing images \cite{abramoff2018pivotal,yang2018generalizing} caused by different sites, scanner vendors, imaging protocols, etc. \par

\begin{figure}[tb!]
	\centering
	\includegraphics[width=1\textwidth]{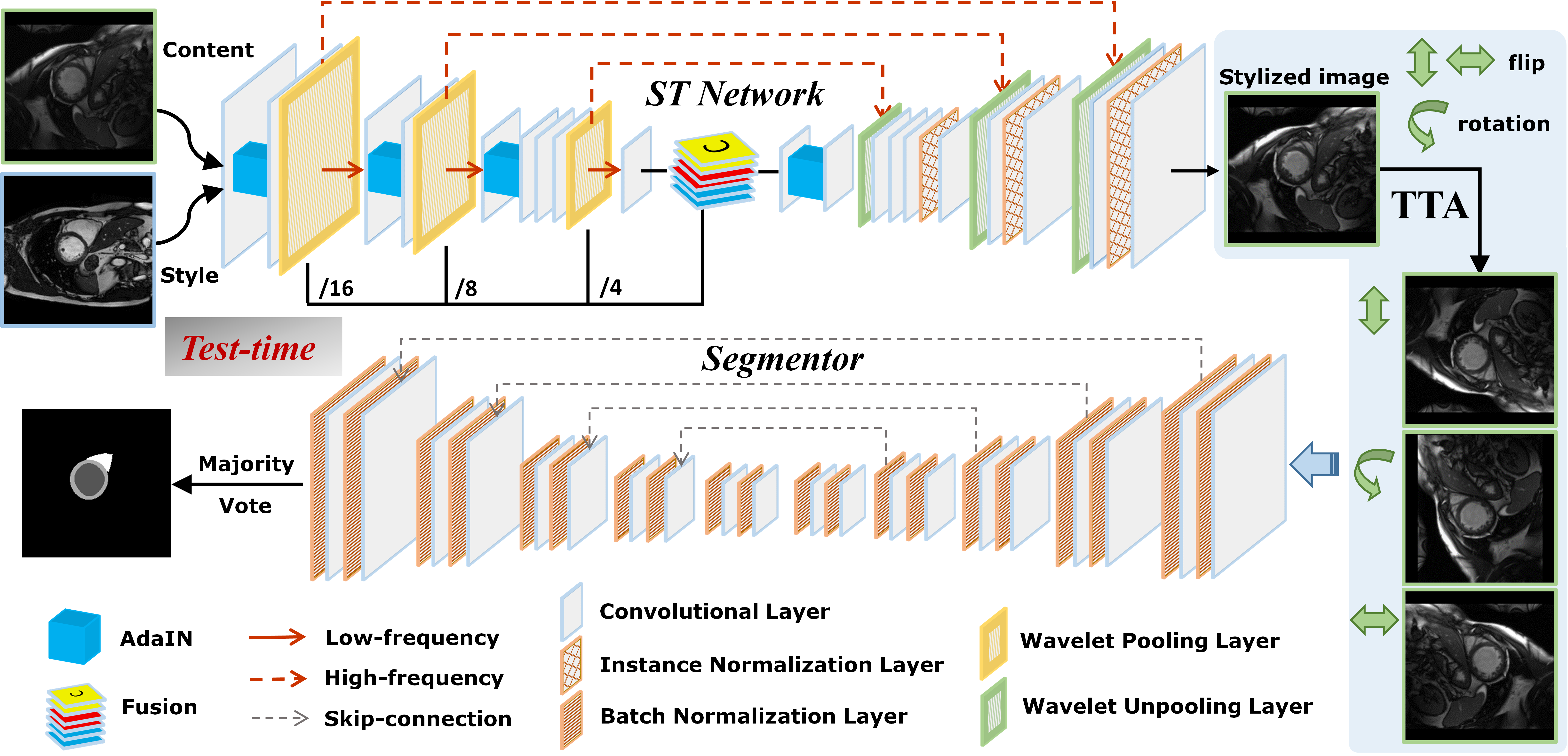}
	\caption{Schematic view of our proposed framework.}
	\label{framework}
\end{figure}

To mitigate the performance degradation, one straightforward choice is data augmentation \cite{perez2017effectiveness,moshkov2020test}. It can help suppress overfitting but cannot guarantee the generalization ability of deep models. Recently, Domain Adaptation (DA) \cite{chen2020unsupervised} and Domain Generalization (DG) \cite{dou2019domain} have been common methods for coping with the appearance shift. As main branches of DA/DG, aligning appearance level or feature level among different domains via adversarial learning were explored. Although DA/DG is attractive, it depends heavily on sufficient data from the target domain or requires enough multiple labeled source data. It is also confined by its domain mapping and may not extend to images from unknown domains. By revisiting the basic definition of appearance shift, style transfer \cite{gatys2016image} (ST) inspires a new and intuitive way for the problem. ST removes appearance shift by rendering the appearance of the content image as the style image \cite{chen2019unsupervised,ma2019neural,liu2020remove}. Compared to DA, ST is independent on the target domain, retraining-free and suitable for images with unknown appearance shifts. In \cite{ma2019neural}, Ma \textit{et al.} made the early attempt to exploit an online ST to reduce the appearance variation for better cardiac MR segmentation. But such optimization-based ST has high latency and restrains real-time applications. Liu \textit{et al.} \cite{liu2020remove} proposed an Adaptive Instance Normalization (AdaIN) \cite{huang2017arbitrary} based ST module for vendor adaption to achieve real-time arbitrary style transfer. However, it directly utilized the pre-trained VGG-16 as the ST backbone, which may be unadaptable for the medical image to retain a more realistic semantic content structure. \par 

In this paper, based on Wavelet Corrected Transfer network (WaveCT) \cite{yoo2019photorealistic} and WaveCT-AIN \cite{liu2020remove}, we propose an improved ST network to generate style-invariant images for removing appearance shift and test-time augmentation to enhance the segmentation results. Our contribution is three-fold. First, inspired by the spirit of universal style transfer, we develop a zero-shot stylization for testing (content) images to generate stylized images that appearance similarity to the source (style) images. Second, we build up a robust model based on the U-Net structure for cardiac segmentation. Our framework is a two-stage system during testing: we utilize the ST network to generate the stylized image, then feed it into the segmentation model. Third, we investigate test-time augmentation to explore transformed versions of the stylized image for inference, followed by inverse transformation and predictions mergence to get the final segmentation result. In particular, we make two experiments to verify our proposed framework. 1) segmentation model trained on the original dataset and 2) segmentation model trained on the style-unified dataset generated by our zero-shot ST network from the original dataset. \par

\section{Methodology}
Fig.\ref{framework} is the schematic view of our proposed method. The universal 2D U-Net and VGG-16 networks serve as the backbones for segmentation and ST, respectively. We first train the segmentation model on the source data, then develop a ST network to generate stylized images that suitable for the segmentation model. Specifically, the proposed framework is a two-stage system for segmenting images with appearance shifts. In the first stage, the testing image is transferred into a stylized image refer to the source image appearance. In the second stage, the segmentation model trained on source data is performed on the stylized image and get the segmentation result. Moreover, we explore transformed versions of the stylized testing image for prediction by using test-time augmentation and then perform a majority vote to obtain the final segmentation result. \par

\subsection{Cardiac Segmentation Network Design}
In this work, we modified the U-Net \cite{ronneberger2015u} as our baseline model, the state-of-the-art 2D semantic segmentation network in medical image analysis. Specifically, we use upsampling instead of deconvolution to avoid the grid effect. The output stride of the network is cut to 16 to reduce overfitting. The Batch Normalization layers are inserted after each convolution layer. The segmentation network aims at predicting four-class pixel-wise probabilistic maps for the three cardiac structures (i.e., LV, MYO, RV) and the background. To train the network, we use a composite segmentation loss function $ L_{seg} $ which consists of two loss terms:
\begin{equation}
\begin{split}
L_{ce} = -\sum_c y^c &log(p^c), \quad
L_{Dice} = \sum_c 1 - \frac{2 \arrowvert X^c \cap Y^c \arrowvert}{\arrowvert X^c \arrowvert + \arrowvert Y^c \arrowvert} \\
&L_{seg} = L_{ce} + \lambda L_{Dice}
\end{split}
\end{equation}
The first term $L_{ce}$ is a categorical cross entropy loss, where $p^c$ denotes the corresponding predicted probability map of different classes. The second term is a Dice loss to measure the similarity between probability map $X^c$ and ground truth $Y^c$. We set $\lambda=0.5$ to balance the contribution of the two losses. \par

\subsection{Zero-shot Style Transfer}
ST enables us to transfer the style of an image called style image to that of an image called the content image, rendering the low-level visual style while preserving its high-level semantic content structure. Inspired by the spirit of ST can remove appearance shift to approach generalize image analysis, we develop a ST network to generate style-invariant images for generalizing segmentation model. Different from the optimization-based or feed-forward approximate stylization, we utilize zero-shot ST to achieve real-time arbitrary stylization without training on any pre-defined styles. \par

To meet the requirement of universal and stable transfer between any content-style pairs, we adopt the WaveCT network recently used in WCT2 \cite{yoo2019photorealistic} and make improvements to preserve image structure details and render the style features. Different from previous online ST methods \cite{ma2019neural} used to remove appearance shift which may distort image details, the WaveCT network replaces vanilla max-pooling/unpooling with the Haar wavelet pooling/unpooling layers that maintain the content structure to the great extent. In particular, WaveCT splits the features into low-frequency and high-frequency components via Haar wavelet pooling, then low-frequency information passes the main network and high-frequency information skips to connect between encoder and decoder. \par

The ST network proposed in this paper is a significant extension of our prior conference paper proposed WaveCT-AIN \cite{liu2020remove}, regarding the following highlighted points. As the ST network depicted in Fig.\ref{framework}, first, we design a multi-scale feature fusion layer after the encoder, in return mitigate the variation of background area without information in the image. Second, we keep the ST module in the encoder and add an extra AdaIN after the feature fusion. Besides, we enhance the style-invariance by introducing the Instance Normalization (IN) layer into the decoder. In this respect, we focus on rendering the style texture representations in the low-level features and preserves its invariance in the high-level patterns. Third, we simplify the case-specific style image selection strategy more concisely and effectively, which is directly considered selecting the reference style image that as close as possible to the mean and standard deviation of the testing image. Especially, the ST network utilizes the pre-trained VGG-16 as backbone while feature fusion layer and IN layers are embedded, thus it needs to be fine-tuned with image reconstruction task. \par

\subsection{Test-time Augmentation}
Data augmentation significantly improves robustness to appearance shift and can be used as a simple strategy for generalizing model performance. Data augmentation at training time has been commonly used to increase the amount of data for improving performance \cite{perez2017effectiveness}. Recent works also demonstrated the usefulness of data augmentation directly at test time, for achieving more robust predictions \cite{moshkov2020test}. For the point of data acquisition, a testing image is only one of many possible observations of the underlying anatomy. Therefore, we explore multiple transformed versions of the testing image for robust segmentation. Test-time augmentation includes four procedures: augmentation, prediction, inverse-augmentation, and merging. We firstly consider different transformations on the testing image. For our case, we have already remove appearance shift through the ST network, thus we apply flip and rotation transformations for stylized cardiac images instead of complicated contrast or brightness change. In particular, we make three different transformations on the stylized testing image and inference each version of the testing image, thus four predictions are obtained by the inverse transformation. Then we perform a majority vote to obtain the final segmentation result, that is, once the pixel is predicted twice or more, it will be regarded as the target area. \par

\section{Experimental Results}
\subsection{Dataset and Implementation Details}
Notably, we make two experiments to verify our proposed framework. \textit{Exp.1}) segmentation model trained on the original training dataset, denoted as \textit{SegO} and \textit{Exp.2}) segmentation model trained on the style-unified dataset generated by our zero-shot ST network from the original dataset, denoted as \textit{SegST}. \par

\textbf{Dataset.} The framework was trained and evaluated on the Multi-Centre, Multi-Vendor \& Multi-Disease Cardiac Image Segmentation Challenge (M\&Ms 2020) dataset \cite{campello2020dataset}. Two subsets of 75 CMR images from vendor A and vendor B (denoted as Ven$_A$ and Ven$_B$) with only ES \& ED annotated are provided as training data, respectively. Additionally, 25 unannotated images are also given from Ven$_C$. However, our methods do not use it because we are concentrate on generalizing the model to other more unknown data, not just Ven$_C$. For evaluation, the results are evaluated on not only 50 new studies from each of Ven$_{A,B,C}$, but also 50 else studies from Ven$_D$.

\begin{table*}[tb!]
	\centering
	\begin{center}
		\caption{Quantitative comparison results of the \textit{Exp.1}.}
		\setlength{\tabcolsep}{1mm}{
			\begin{tabular}{l|cccc|cccc|cccc}
				\toprule[1.5pt]
				\multirow{2}{*}{\textbf{Metrics}} & \multicolumn{4}{c|}{\scriptsize\textbf{SegO}} &
				\multicolumn{4}{c|}{\scriptsize\textbf{STSegO}} &
				\multicolumn{4}{c}{\scriptsize\textbf{STSegO-TTA}} \\
				\cline{2-13} 
				& \tiny\textbf{Ven$_A$} & \tiny\textbf{Ven$_B$} & \tiny\textbf{Ven$_C$} & \tiny\textbf{Ven$_D$} 
				& \tiny\textbf{Ven$_A$} & \tiny\textbf{Ven$_B$} & \tiny\textbf{Ven$_C$} & \tiny\textbf{Ven$_D$} 
				& \tiny\textbf{Ven$_A$} & \tiny\textbf{Ven$_B$} & \tiny\textbf{Ven$_C$} & \tiny\textbf{Ven$_D$} 
				\\
				\hline
				\scriptsize{\textbf{Dice$_{AVG}$}} & 
				\scriptsize{80.72} & \scriptsize{86.82} & \scriptsize{81.20} & \scriptsize{62.23} & 
				\scriptsize{82.92} & \scriptsize{\textbf{89.72}} & \scriptsize{85.27} & \scriptsize{64.87} & \scriptsize{\textbf{84.70}} & \scriptsize{89.57} & \scriptsize{\textbf{85.56}} & \scriptsize{\textbf{68.01}} \\
				\scriptsize{\textbf{Jac$_{AVG}$}} & 
				\scriptsize{68.93} & \scriptsize{77.98} & \scriptsize{69.53} & \scriptsize{49.81} & 
				\scriptsize{71.82} & \scriptsize{\textbf{82.04}} & \scriptsize{75.01} & \scriptsize{53.49} & \scriptsize{\textbf{74.29}} & \scriptsize{81.98} & \scriptsize{\textbf{75.45}} & \scriptsize{\textbf{55.80}} \\
				\scriptsize{\textbf{HDB$_{AVG}$}} &
				\scriptsize{19.52} & \scriptsize{14.24} & \scriptsize{18.30} & \scriptsize{44.88} & 
				\scriptsize{17.54} & \scriptsize{9.93} & \scriptsize{13.27} & \scriptsize{44.53} & \scriptsize{\textbf{14.88}} & \scriptsize{\textbf{9.04}} & \scriptsize{\textbf{12.94}} & \scriptsize{\textbf{38.46}} \\
				\scriptsize{\textbf{ASSD$_{AVG}$}} & 
				\scriptsize{1.96} & \scriptsize{0.89} & \scriptsize{1.84} & \scriptsize{12.44} & 
				\scriptsize{1.85} & \scriptsize{0.65} & \scriptsize{1.46} & \scriptsize{13.17} & \scriptsize{\textbf{1.51}} & \scriptsize{\textbf{0.65}} & \scriptsize{\textbf{1.44}} & \scriptsize{\textbf{7.38}} \\
				\hline
				\scriptsize{\textbf{Dice$_{LV}$}} & 
				\scriptsize{89.14} & \scriptsize{93.07} & \scriptsize{85.12} & \scriptsize{72.97} & 
				\scriptsize{88.71} & \scriptsize{\textbf{92.90}} & \scriptsize{\textbf{86.46}} & \scriptsize{\textbf{78.21}} & \scriptsize{\textbf{89.75}} & \scriptsize{92.45} & \scriptsize{86.39} & \scriptsize{74.91} \\
				\scriptsize{\textbf{Jac$_{LV}$}} & 
				\scriptsize{81.01} & \scriptsize{87.56} & \scriptsize{75.56} & \scriptsize{62.83} & 
				\scriptsize{80.27} & \scriptsize{\textbf{87.39}} & \scriptsize{\textbf{77.34}} & \scriptsize{\textbf{68.09}} & \scriptsize{\textbf{81.95}} & \scriptsize{86.92} & \scriptsize{77.26} & \scriptsize{64.94} \\
				\scriptsize{\textbf{HDB$_{LV}$}} & 
				\scriptsize{13.58} & \scriptsize{9.55} & \scriptsize{13.38} & \scriptsize{33.68} & 
				\scriptsize{13.28} & \scriptsize{6.72} & \scriptsize{12.05} & \scriptsize{\textbf{28.23}} & \scriptsize{\textbf{11.59}} & \scriptsize{\textbf{6.03}} & \scriptsize{\textbf{10.90}} & \scriptsize{31.42} \\
				\scriptsize{\textbf{ASSD$_{LV}$}} & 
				\scriptsize{1.53} & \scriptsize{0.61} & \scriptsize{1.91} & \scriptsize{10.86} & 
				\scriptsize{1.54} & \scriptsize{\textbf{0.63}} & \scriptsize{\textbf{1.74}} & \scriptsize{\textbf{5.76}} & \scriptsize{\textbf{1.34}} & \scriptsize{0.68} & \scriptsize{1.76} & \scriptsize{8.28} \\
				\hline
				\scriptsize{\textbf{Dice$_{MYO}$}} & 
				\scriptsize{72.71} & \scriptsize{76.63} & \scriptsize{73.60} & \scriptsize{51.76} & 
				\scriptsize{81.84} & \scriptsize{\textbf{85.69}} & \scriptsize{83.83} & \scriptsize{60.76} & \scriptsize{\textbf{83.13}} & \scriptsize{85.68} & \scriptsize{\textbf{83.92}} & \scriptsize{\textbf{63.19}} \\
				\scriptsize{\textbf{Jac$_{MYO}$}} & 
				\scriptsize{57.47} & \scriptsize{62.64} & \scriptsize{58.72} & \scriptsize{36.92} & 
				\scriptsize{69.49} & \scriptsize{75.19} & \scriptsize{72.52} & \scriptsize{47.01} & \scriptsize{\textbf{71.32}} & \scriptsize{\textbf{75.21}} & \scriptsize{\textbf{72.63}} & \scriptsize{\textbf{48.53}} \\
				\scriptsize{\textbf{HDB$_{MYO}$}} & 
				\scriptsize{20.27} & \scriptsize{21.56} & \scriptsize{19.15} & \scriptsize{34.00} & 
				\scriptsize{16.54} & \scriptsize{11.36} & \scriptsize{\textbf{14.77}} & \scriptsize{\textbf{27.22}} & \scriptsize{\textbf{14.02}} & \scriptsize{\textbf{10.74}} & \scriptsize{14.85} & \scriptsize{30.40} \\
				\scriptsize{\textbf{ASSD$_{MYO}$}} & 
				\scriptsize{1.92} & \scriptsize{1.35} & \scriptsize{2.00} & \scriptsize{7.13} & 
				\scriptsize{1.29} & \scriptsize{0.59} & \scriptsize{1.29} & \scriptsize{\textbf{2.67}} & \scriptsize{\textbf{1.09}} & \scriptsize{\textbf{0.55}} & \scriptsize{\textbf{1.28}} & \scriptsize{3.04} \\
				\hline
				\scriptsize{\textbf{Dice$_{RV}$}} & 
				\scriptsize{80.31} & \scriptsize{90.77} & \scriptsize{84.88} & \scriptsize{61.96} & 
				\scriptsize{78.21} & \scriptsize{90.58} & \scriptsize{85.51} & \scriptsize{55.66} & \scriptsize{\textbf{81.22}} & \scriptsize{\textbf{90.60}} & \scriptsize{\textbf{86.39}} & \scriptsize{\textbf{65.94}} \\
				\scriptsize{\textbf{Jac$_{RV}$}} & 
				\scriptsize{68.32} & \scriptsize{83.74} & \scriptsize{74.30} & \scriptsize{49.69} & 
				\scriptsize{65.71} & \scriptsize{83.55} & \scriptsize{75.18} & \scriptsize{45.38} & \scriptsize{\textbf{69.60}} & \scriptsize{\textbf{83.79}} & \scriptsize{\textbf{76.47}} & \scriptsize{\textbf{53.94}} \\
				\scriptsize{\textbf{HDB$_{RV}$}} & 
				\scriptsize{24.72} & \scriptsize{11.61} & \scriptsize{22.38} & \scriptsize{66.95} & 
				\scriptsize{22.79} & \scriptsize{11.71} & \scriptsize{\textbf{12.98}} & \scriptsize{78.14} & \scriptsize{\textbf{19.03}} & \scriptsize{\textbf{10.35}} & \scriptsize{13.07} & \scriptsize{\textbf{53.55}} \\
				\scriptsize{\textbf{ASSD$_{RV}$}} & 
				\scriptsize{2.43} & \scriptsize{0.70} & \scriptsize{1.61} & \scriptsize{19.34} & 
				\scriptsize{2.72} & \scriptsize{0.74} & \scriptsize{1.35} & \scriptsize{31.08} & \scriptsize{\textbf{2.09}} & \scriptsize{\textbf{0.72}} & \scriptsize{\textbf{1.27}} & \scriptsize{\textbf{10.83}} \\
				\bottomrule[1.5pt]
		\end{tabular}}
		\label{tab1}
	\end{center}
\end{table*}

\textbf{Implementation Details.} We obtain 3284 training slices along the anatomical plane from the ES \& ED images of Ven$_A$ and Ven$_B$. All slices are resized to 256 $\times$ 256. For training the \textit{SegO}, we apply elastic deformations to the available training slices (i.e., random expand, flip, rotation, mirror, contrast change, and brightness change). Whereas the training of \textit{SegST} is not necessary to make contrast or brightness change because its training data already have a particular appearance distribution. We use 3000 cardiac slices to fine-tune the ST Network as the extra feature fusion layer and IN layers are embedded into the pre-trained VGG-16. For the segmentation network, it was trained for 60k iterations with a batch size of 24 and was optimized using the composite loss $L_{seg}$ where Adam optimizer with a learning rate of $10^{-3}$ initially then decreased to $10^{-5}$. We implement all experiments with PyTorch on two GeForce$^{\textregistered}$ RTX 2080 Ti GPU.

\subsection{Quantitative and Qualitative Evaluation}
\textbf{Metrics.} To evaluate the accuracy of segmentation performance, we adopt in total 4 indicators including the Dice similarity index (Dice, \%), Jaccard similarity index (Jac, \%), Hausdorff Distance of Boundaries (HDB, pixel), and Average Symmetric Surface Distance (ASSD, pixel). For ease of comparison, we calculate the average (AVG) per indicator over the three structures (LV, MYO, RV).

\begin{figure}[tb!]
	\centering
	\includegraphics[width=0.95\textwidth]{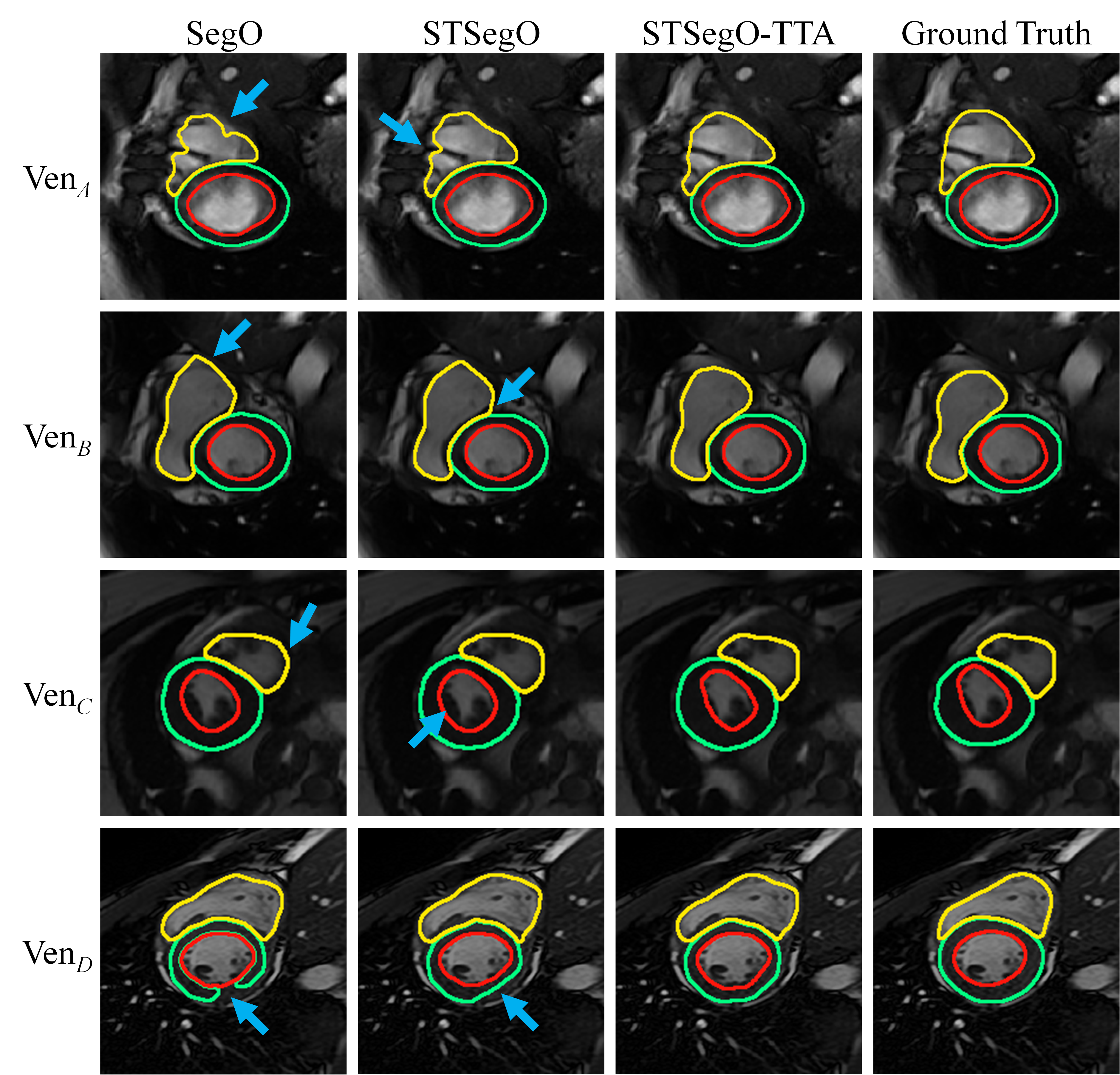}
	\caption{Visualization of the 2D segmentation results of our proposed methods. Green, red and yellow curves represent LV, MYO and RV, respectively. The performance is gradually improved from left to right methods, especially in the boundaries of RV.}
	\label{vis_slice}
\end{figure}

\subsubsection{Quantitative Results of the \textit{Exp.1}.}
\begin{figure}[tb!]
	\centering
	\includegraphics[width=0.95\textwidth]{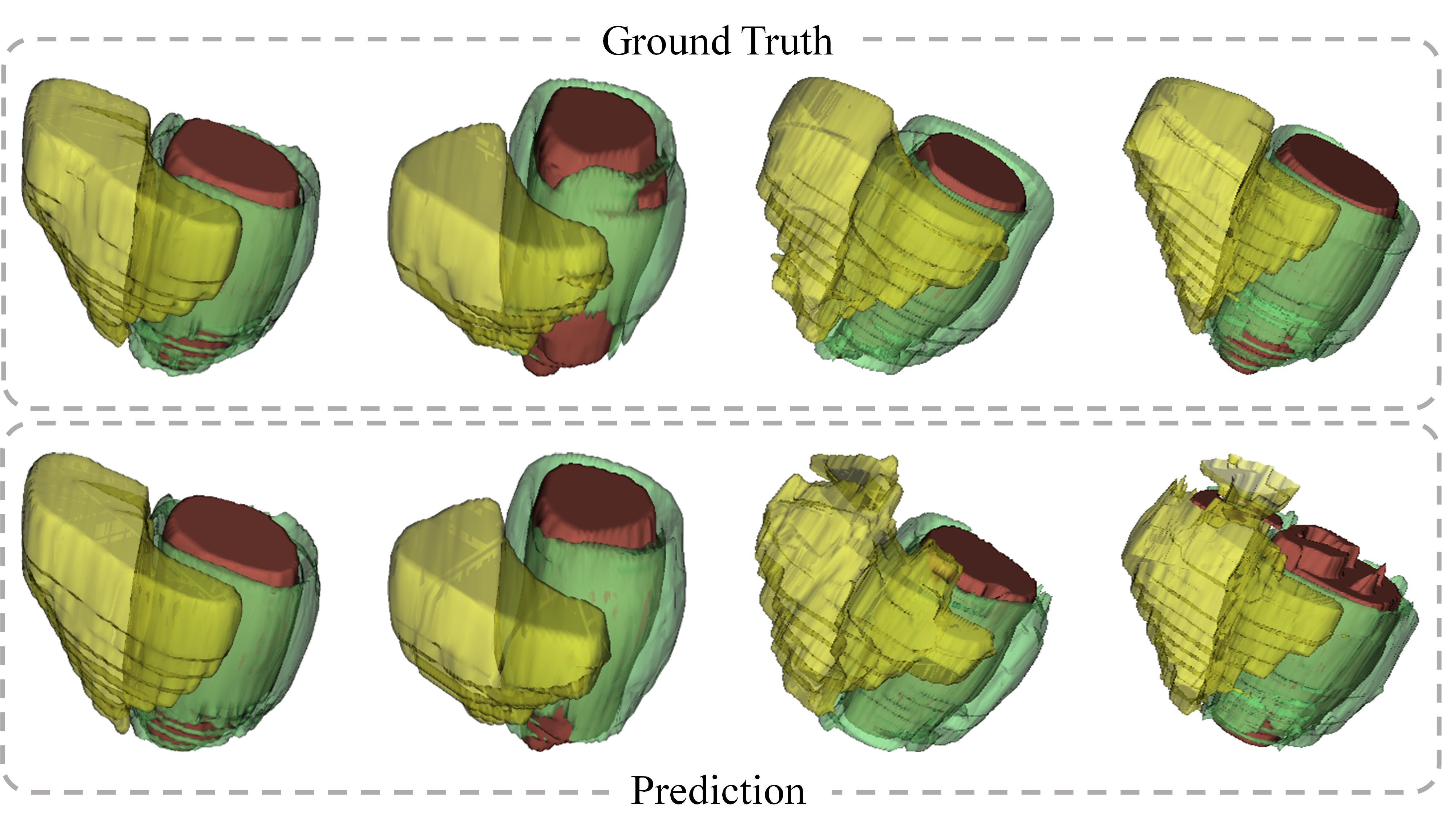}
	\caption{Visualization of our better 3D segmentation results. From left to right are cases from Ven$_B$, Ven$_C$, Ven$_D$ and Ven$_D$, respectively. Green, red and yellow areas represent LV, MYO and RV, respectively.}
	\label{vis_volume}
\end{figure}

We first train the segmentation model \textit{SegO} based on the available labeled images from Ven$_A$ and Ven$_B$. Then we set up a style image library from the training data, which contains 221 slices from the top-20 images via computing the Dice index. Therefore, the testing (content) slice selects the reference style slice from the library through our simple style selection strategy. Subsequently, the content-style pairs feed into our zero-shot ST network to generate the stylized slice for segmentation. This two-stage system is denoted as \textit{STSegO}. Likewise, \textit{STSegO} with test-time augmentation is denoted as \textit{STSegO-TTA}. Table.\ref{tab1} shows the quantitative results based on the 200 test cases correspond to four different vendors. We compare three versions of our proposed \textit{SegO}, \textit{STSegO} and \textit{STSegO-TTA}, respectively. The numbers in bold indicate the best results of multiple vendors among different methods. Both \textit{STSegO} and \textit{STSegO-TTA} get consistent improvements over the pure \textit{SegO}, in which the best results are achieved by the \textit{STSegO-TTA}. It almost improves the Dice index by 5 percent and the Jaccard index by 6 percent on average for each vendor. The HDB and ASSD also improve about 5.5 pixels and 1.5 pixels, respectively. Obviously, the segmentation model can be well generalized to the images of Ven$_C$, but the performance on the Ven$_D$ shows relatively poor. This may be due to the large difference in the data distribution and anatomical structure between images of Ven$_D$ and the source data. Fig. \ref{vis_slice} visualizes the 2D segmentation results of our proposed methods on unseen cases from four vendors. \textit{STSegO-TTA} produces more anatomically plausible results on the images. Fig. \ref{vis_volume} visualizes the 3D segmentation results of the \textit{STSegO-TTA} on four unseen cases. \par

\begin{table*}[tb!]
	\centering
	\begin{center}
		\caption{Quantitative comparison results of the \textit{Exp.2}.}
		\setlength{\tabcolsep}{1mm}{
			\begin{tabular}{l|cccc|cccc}
				\toprule[1.5pt]
				\multirow{2}{*}{\scriptsize\textbf{Metrics}} & \multicolumn{4}{c|}{\scriptsize\textbf{SegST}} &
				\multicolumn{4}{c}{\scriptsize\textbf{SegST-TTA}} \\
				\cline{2-9} 
				& \tiny\textbf{Ven$_A$} & \tiny\textbf{Ven$_B$} & \tiny\textbf{Ven$_C$} & \tiny\textbf{Ven$_D$} 
				& \tiny\textbf{Ven$_A$} & \tiny\textbf{Ven$_B$} & \tiny\textbf{Ven$_C$} & \tiny\textbf{Ven$_D$} 
				\\
				\hline
				\scriptsize{\textbf{Dice$_{AVG}$}} & \scriptsize{84.19} & \scriptsize{89.63} & \scriptsize{\textbf{85.74}} & \scriptsize{53.84} & 
				\scriptsize{\textbf{85.99}} & \scriptsize{\textbf{90.28}} & \scriptsize{85.23} & 
				\scriptsize{\textbf{58.86}} \\
				
				\scriptsize{\textbf{Jac$_{AVG}$}} & \scriptsize{73.31} & \scriptsize{82.06} & \scriptsize{\textbf{75.71}} & \scriptsize{42.64} & 
				\scriptsize{\textbf{76.03}} & \scriptsize{\textbf{82.74}} & \scriptsize{74.96} & 
				\scriptsize{\textbf{47.90}} \\
				
				\scriptsize{\textbf{HDB$_{AVG}$}} & \scriptsize{16.89} & \scriptsize{15.46} & \scriptsize{20.22} & \scriptsize{\textbf{35.56}} & 
				\scriptsize{\textbf{13.62}} & \scriptsize{\textbf{9.28}} & \scriptsize{\textbf{13.39}} & 
				\scriptsize{47.86}  \\
				
				\scriptsize{\textbf{ASSD$_{AVG}$}} & \scriptsize{1.71} & \scriptsize{0.73} & \scriptsize{1.44} & \scriptsize{\textbf{14.55}} & 
				\scriptsize{\textbf{1.34}} & \scriptsize{\textbf{0.58}} & \scriptsize{\textbf{1.37}} & 
				\scriptsize{17.52}  \\

				\hline
				\scriptsize{\textbf{Dice$_{LV}$}} & \scriptsize{87.82} & \scriptsize{92.54} & \scriptsize{87.04} & \scriptsize{63.71} & 
				\scriptsize{\textbf{89.64}} & \scriptsize{\textbf{93.83}} & \scriptsize{\textbf{87.50}} & \scriptsize{\textbf{68.34}}   \\
				
				\scriptsize{\textbf{Jac$_{LV}$}} & \scriptsize{78.84} & \scriptsize{86.95} & \scriptsize{78.23} & \scriptsize{52.92} & 
				\scriptsize{\textbf{81.85}} & \scriptsize{\textbf{88.68}} & \scriptsize{\textbf{78.87}} & 
				\scriptsize{\textbf{58.54}}  \\
				
				\scriptsize{\textbf{HDB$_{LV}$}} & \scriptsize{16.78} & \scriptsize{14.45} & \scriptsize{17.34} & \scriptsize{\textbf{48.96}} & 
				\scriptsize{\textbf{10.96}} & \scriptsize{\textbf{6.22}} & \scriptsize{\textbf{11.52}} & 
				\scriptsize{49.84}  \\
				
				\scriptsize{\textbf{ASSD$_{LV}$}} & \scriptsize{1.90} & \scriptsize{0.76} & \scriptsize{1.74} & \scriptsize{\textbf{20.87}} & 
				\scriptsize{\textbf{1.33}} & \scriptsize{\textbf{0.47}} & \scriptsize{\textbf{1.49}} & 
				\scriptsize{21.27}  \\

				\hline
				\scriptsize{\textbf{Dice$_{MYO}$}} & \scriptsize{81.04} & \scriptsize{\textbf{86.61}} & \scriptsize{\textbf{84.81}} & \scriptsize{53.49} & 
				\scriptsize{\textbf{82.92}} & \scriptsize{86.54} & \scriptsize{84.70} & 
				\scriptsize{\textbf{56.97}}  \\
				
				\scriptsize{\textbf{Jac$_{MYO}$}} & \scriptsize{68.37} & \scriptsize{\textbf{76.81}} & \scriptsize{\textbf{73.96}} & \scriptsize{40.61} & 
				\scriptsize{\textbf{71.00}} & \scriptsize{76.44} & \scriptsize{73.74} & 
				\scriptsize{\textbf{44.49}}  \\
				
				\scriptsize{\textbf{HDB$_{MYO}$}} & \scriptsize{16.42} & \scriptsize{20.60} & \scriptsize{22.42} & \scriptsize{\textbf{24.78}} & 
				\scriptsize{\textbf{13.46}} & \scriptsize{\textbf{10.11}} & \scriptsize{\textbf{14.08}} & 
				\scriptsize{30.12}  \\
				
				\scriptsize{\textbf{ASSD$_{MYO}$}} & \scriptsize{1.39} & \scriptsize{0.59} & \scriptsize{1.28} & \scriptsize{12.26} & 
				\scriptsize{\textbf{1.07}} & \scriptsize{\textbf{0.49}} & \scriptsize{\textbf{1.01}} & 
				\scriptsize{\textbf{7.68}} \\
				
				\hline
				\scriptsize{\textbf{Dice$_{RV}$}} & \scriptsize{83.70} & \scriptsize{89.73} & \scriptsize{\textbf{85.37}} & \scriptsize{44.31} & 
				\scriptsize{\textbf{85.41}} & \scriptsize{\textbf{90.46}} & \scriptsize{83.49} & 
				\scriptsize{\textbf{51.26}}  \\
				
				\scriptsize{\textbf{Jac$_{RV}$}} & \scriptsize{72.71} & \scriptsize{82.42} & \scriptsize{\textbf{74.94}} & \scriptsize{34.40} & 
				\scriptsize{\textbf{75.25}} & \scriptsize{\textbf{83.09}} & \scriptsize{72.28} & 
				\scriptsize{\textbf{40.66}}  \\
				
				\scriptsize{\textbf{HDB$_{RV}$}} & \scriptsize{17.45} & \scriptsize{\textbf{11.32}} & \scriptsize{20.90} & \scriptsize{\textbf{32.93}} & 
				\scriptsize{\textbf{16.45}} & \scriptsize{11.49} & \scriptsize{\textbf{14.57}} & 
				\scriptsize{63.62}  \\
				
				\scriptsize{\textbf{ASSD$_{RV}$}} & \scriptsize{1.85} & \scriptsize{0.84} & \scriptsize{\textbf{1.30}} & \scriptsize{\textbf{10.52}} & 
				\scriptsize{\textbf{1.64}} & \scriptsize{\textbf{0.79}} & \scriptsize{1.62} & 
				\scriptsize{23.62}  \\
				
				\bottomrule[1.5pt]
		\end{tabular}}
		\label{tab2}
	\end{center}
\end{table*}

\subsubsection{Quantitative Results of the \textit{Exp.2}.}
Different from the \textit{SegO} trained on the original dataset, \textit{SegST} utilized the style-unified dataset for training, which is generated by our zero-shot ST network from the original dataset. Notably, we randomly select a slice serve as the style slice to generate stylized data, thus the style-unified training data has a particular appearance distribution. Consequently, the ST network only takes over this style slice to achieve stylization during testing. Likewise, \textit{SegST} with test-time augmentation is denoted as \textit{SegST-TTA}. As can be seen from Table.\ref{tab2}, \textit{STSeg-TTA} shows improvements over \textit{STSeg}, the Dice index and Jaccard index are both raised about 2 percent on average for each vendor, and the HDB is improved about 1 pixel. However, the performance of Ven$_D$ shows worse compared with \textit{Exp.1}, which may be caused by the target style slice is not suitable for the images of Ven$_D$. Thus it is crucial to choose a universal style slice to generate the style-unified dataset, which will be our further study.

\section{Conclusion}
In this paper, we proposed a zero-shot ST network to generate style-invariant images for removing appearance shift and test-time augmentation to enhance the segmentation results. By investigating the two experiments \textit{Exp.1} and \textit{Exp.2}, we showed that \textit{SegO} and \textit{STSeg} with their variants present promising performance in segmenting cardiac images across the multi-vendor and multi-cencre dataset.

\section{Acknowledgement}
\label{sec:Acknowledgement}
This work was supported by the grant from National Key R\&D Program of China (No. 2019YFC0118300), Shenzhen Peacock Plan (No. KQTD2016053112051497, KQJSCX20180328095606003) and National Natural Science Foundation of China (No.61801296).

\bibliographystyle{splncs04}
\bibliography{mnms-043}

\end{document}